\documentclass{appolb}
\usepackage{epsfig}
\begin{document}
%\eqsec  % uncomment this line to get equations numbered by (sec.num)
\title{Large rapidity gaps survival probabilities at LHC
\thanks{Presented at ISMD07}%
}
\author{Rohini M. Godbole
\address{Centre for High Energy Physics, Indian Institute of Science,Bangalore,
560012, India}
\and
Agnes Grau
\address{Departamento de F\'\i sica Te\'orica y del Cosmos, Universidad de 
Granada, Spain}
\and
Giulia Pancheri
\address{INFN Frascati National Laboratories, P.O. Box 13, Frascati, I00044  
Italy}
\and
Yogendra N. Srivastava
\address{Physics Department and INFN, University of Perugia,
             Perugia, Italy}
}
\maketitle
\begin{abstract}
We 
%apply a recently developed model for minimum bias collisions  to 
calculate the  probability of large rapidity gaps in high energy hadronic 
collisions
%. The predictions of this 
using a model
%, which is 
based on QCD mini-jets and soft gluon emission down into the infrared 
region.
%, are compared
Comparing  with other models 
%and 
we find a remarkable agreement among most predictions.
\end{abstract}
  
\section{Introduction}
%The existence of large rapidity gaps  in hadronic collisions was
%proposed 
The possibility of the existence of large rapidity gaps in hadronic
collisions was suggested a number of years ago \cite{bjorken,khoze}.
% and observed in high energy collisions\cite{CDF}. 
In \cite{bjorken}, the occurrence   of rapidity regions deprived of soft 
hadronic debris was proposed as a means to search for Higgs bosons produced 
by interactions between colour singlet particles, e.g. $W,Z$ bosons  
emitted by initial state quarks, but with a warning
% Bjorken suggested to use the rapidity gaps to sharpen the search for the 
%Higgs boson, but warned 
against the possibility of soft collisions which would populate these 
regions. For an estimate, an expression 
%This can be done calculating 
for the Large Rapidity Gaps Survival Probability (LRGSP)
% for which  the following expression 
was proposed, namely
%The gap survival%probability
%~\cite{Bjorken:1991xr,Bjorken:1992er} %is thus given by
\begin{equation}< |S | ^2 > = \frac{\int d^2 {\bf b}\  
A^{AB}( b ,s) | S( b) |^2 \sigma_H(b,s)}{\int d^2 {\bf b}\  
A^{AB}( b,s)\sigma_H(b,s)}\label{sob}
\end{equation}
where $| S( b) |^2$ is the probability in impact parameter space 
that two hadrons A,B got through each other without detectable inelastic 
interactions,  
%until an impact parameter $\bf b$;  
$A^{AB}(b,s)$ is the $b-$distribution for  
%interactions in which only 
collisions involved in interactions in which only low
$p_t$ particle emission can take place, and  
$\sigma_H(b,s)$ is the cross-section for producing, say, a Higgs  boson, in
such hadronically deprived  silent configuration.
%in the interaction between the two hadrons A and B, 
%when they collide at an 
%impact parameter distance $\bf b$. 
To use the above equation, one needs to estimate the probability of 
$ not$ having inelastic collisions, 
$|S(b)|^2=P_{no-inel}(b)$ and the distribution for 
low-$p_t$ interactions, namely the probability $A^{AB}(b,s)$
%\equiv A^{soft}(b,s)$ 
to find  those collisions for which  two hadrons A and B  at 
distance ${\bf b}$   will not undergo large $p_t$ collisions. 
Introducing for the scattered partons a cut-off $p_{tmin}$, 
above which the scattering process can be described by perturbative QCD, 
we need to estimate the $b-$distribution of all the collisions 
 with $p_t<p_{tmin}$.
%{\bf Question:  this terminology should include also hard collinear, 
%or not? Any help here?}
In order to calculate  $P_{no-inel}$, one can approximate the bulk of 
hadron-hadron collisions with a sum of Poisson distributions for 
$k$ independent collisions distributed around an average  
${\bar n}\equiv n(b,s)$, namely 
\begin{equation}
P_{no-inel}=
1-\sum_k \Pi\{k,{\bar n}\} =
1-\sum_{k=1}^\infty {{{\bar n}^k e^{-{\bar n}}}\over{k!}}=e^{-n(b,s)}
\end{equation}
which also leads to the following expressions for the total and inelastic 
cross-sections
\begin{equation}\sigma_{inel}^{AB}=\int d^2{\bf b}[1-e^{-n(b,s)}]
\end{equation}
and 
\begin{equation}
\sigma_{tot}^{AB}=2\int d^2{\bf b}[1-e^{-\frac {n(b,s)}{2}}
\cos{\Re e\chi(b,s)}]
\end{equation}
where $n(b,s)/2$ can be  identified with the imaginary part of the eikonal 
function $\chi(b,s)$. Approximating $\Re e\chi(b,s)=0$,
 allows a simple way of calculating 
$\sigma_{tot}$ if one has a model for $n(b,s)$. 
A possible strategy to calculate the LRGSP is then to build a model for the 
total cross-section and then insert the relevant $b-$distributions in  
Eq.(\ref{sob}).
% a model for the impact parameter distribution of hadrons 
%and the possibility to distinguish between hard and soft collisions. 
%Such distributions are present 
%in calculations of total cross-section based on the eikonal
%representation, albeit in different 
%approximations. 
The basic quantity to evaluate is thus
\begin{equation}
n(b,s)=n^{NP}(b,s)+n^{hard}(b,s)=A^{NP}(b,s)
\sigma_{NP}(s)+A^{hard}(b,s)\sigma_{hard}(s)
\end{equation}
where we have split the average number of collisions between those with 
outgoing partons below (NP) and above (hard) the cut-off $p_{tmin}$. 
In the next section we shall describe our model 
\cite{ourplb,ourprd} for total cross-section, comparing its results to 
other models. Then,  we shall use the $b-$distributions from our model to 
evaluate the LRGSP at LHC, again comparing it to different model results.

\section{The Eikonal Mini-jet Model  for total cross-section}
To build a realistic model for $\sigma_{tot}$, one needs to understand what 
makes the cross-section rise 
%and what makes it rise 
with a slope compatible 
with the limits imposed by the Froissart theorem, 
namely $\sigma_{tot} \le \log^2{s}$. In our model, 
the rise in $\sigma_{tot}$ is driven by the rise 
of low-x (perturbative) gluon-gluon interactions, 
while the saturation imposed by the Froissart bound 
comes from initial state emission of infrared gluons  
which temper the too fast rise of the minijet cross-section.  
The rise is calculated using perturbative QCD for collisions producing 
partons with  $p_t>p_{tmin}\approx 1\div 2 GeV$, using 
%\end{document}
%namely 
%we put 
%\begin{equation}\sigma^{AB}_{\rm jet} (s;p_{tmin}) = 
%\int_{p_{tmin}}^{\sqrt{s}/2} d p_t \int_{4p_t^2/s}^1 d x_1 
%\int_{4 p_t^2/(x_1 s)}^1 d x_2 \sum_{i,j,k,l}f_{i|A}(x_1,p_t^2) 
%f_{j|B}(x_2, p_t^2)~~ \frac { d \hat{\sigma}_{ij}^{ kl}(\hat{s})} 
%{d p_t}.
%\label{sigjet}\end{equation} 
%Here  subscripts $A$ and $B$ denote the colliding particles 
%($p$ and/or  $\bar p$), $i, \ j, \ k, \ l$ the partons and $x_1,x_2$ the 
%fractions of the parent particle momentum carried by the parton. 
%$\hat{s} =x_1 x_2 s$  and $\hat{ \sigma}$ are 
hard parton scattering cross-sections,  
%$f_{i|A}(x,p_t)$ are the 
and the experimentally measured and DGLAP evolved parton densities (PDF's) 
in the scattering hadrons. Thus parton densities and the elements of 
perturbative QCD are the only input needed for the calculation of  
%\end{document}
$\sigma_{hard}$.
%\end{document}
%\sigma_{\rm jet}^{\rm AB}$. 
The rate of rise of this cross-section with energy is determined by 
$p_{tmin}$ and the low-x behaviour of the parton densities. 
As noted before, the rise with energy of 
%this rise 
the cross-section obtained with this
is much steeper than 
that consistent with the Froissart bound, but in our model this rise is 
tempered  by soft gluon emission. The saturation mechanism takes place 
through the $b$-distribution obtained from  the Fourier transform of the 
resummed infrared gluon distribution. This distribution is energy dependent 
and given by \cite{ourcorsetti}
\begin{equation} 
A(b,s)=A_0 \int
 d^2 {\bf K}_t e^{-i{\bf K}_t\cdot {\bf b}}\Pi({\bf K}_t)=\frac 
{ e^{-h(b,q_{max})}}{\int d^2{\bf b}  e^{-h(b,q_{max})}}\equiv A_{BN} 
(b,q_{max})\label{adb}
\end{equation}
where the function  $h(b,q_{max}) $ is obtained through summing soft 
gluons \cite{ourcorsetti} and requires integration of soft gluon momenta 
from zero to $q_{max}$ the maximum transverse momentum allowed by 
kinematics to single soft gluons. The saturation of the Froissart bound 
is due to the increasing acollinearity of ``hard'' partons produced by 
initial state soft gluon emission. The single soft gluon distribution 
needed for the calculation of $h(b,q_{max})$ requires using  
infra-red $k_t$ gluons and different models with a frozen or singular 
$\alpha_{strong}(k_t)$ produce different saturation effects. We have shown 
\cite{oldPRD} that the frozen model  is inadequate to quench the rise 
due to minijets, since we see  that the early  rise in 
proton-antiproton collisions requires  minijets with $p_{tmin}\approx 
1 \ GeV$, but then the total cross-section rises too much. 
Put differently, but equivalently, 
raising $p_{tmin}$ to fit higher energy values of the cross-section, 
say  at the Tevatron,  would  require $p_{tmin}\approx 2\ GeV$ but 
 miss the early rise. Instead, we find that a singular 
$\alpha_s$ produces an adequate $s-$dependent saturation effect. 
Singular expressions for $\alpha_s$ are discussed in the literature, 
in particular for quarkonium phenomenology \cite{Yndurain}. 
Our choice is a singular, but integrable expression for the strong 
coupling constant in the infrared region, namely
\begin{equation}
\alpha_s(k_t)\approx 
{{12 \pi}\over{33-2N_f}}({{\Lambda_{QCD}}\over{k_t}})^{2p}
\ \ \ \ \ \ \ k_t \rightarrow 0
\end{equation}
where the scale factor is chosen to allow a smooth interpolation to the 
asymptotic freedom expression for $\alpha_s$, namely we 
choose
\begin{equation}
\alpha_s(k_t)={{12 \pi}\over{33-2N_f}}
 {{p}\over{\log(1+p({{k_t}\over{\Lambda_{QCD}}})^{2p})}}
\end{equation}
The singularity in the infrared is regulated by the parameter $p$, 
which has to be $<1$ for the integral in 
$h(b,q_{max})$ to converge. 
%In a recently developed model \cite{ourplb} 
%for $\sigma_{tot}$, we have used the $b$-distribution derived from soft 
%gluon emission \cite{ourcorsetti}. 
The next input for 
phenomenological tests of our model is the number of 
non-perturbative (NP) collisions. 
We approximate it as\begin{equation}n^{NP}=
A_{BN}^{NP}\sigma_0(1+{{2\epsilon}\over{\sqrt{s}}})
\end{equation}
with $\epsilon=0,1$ for the process $pp$ or $p{\bar p}$. 
We choose a constant $\sigma_0\approx 48 \ mb$ and use 
for $A_{BN}^{NP}$ the same model as for the hard collisions, 
%but for the fact that $q_{max}$ is taken not be larger 
but we restrict $q_{max}$ to be no larger 
than $\approx 20\% p_{tmin}$, since these collisions are 
limited to $p_t<p_{tmin}$. The same function $A_{BN}^{NP}$ is 
then used for the LRGSP calculation, as discussed in the next section 
where both the  estimated $\sigma_{total}$ as well as the LRGSP will be 
presented and compared with other models.
\section{Total cross-sections and survival probability}
\begin{figure}
\begin{center}
\includegraphics[scale=0.3]{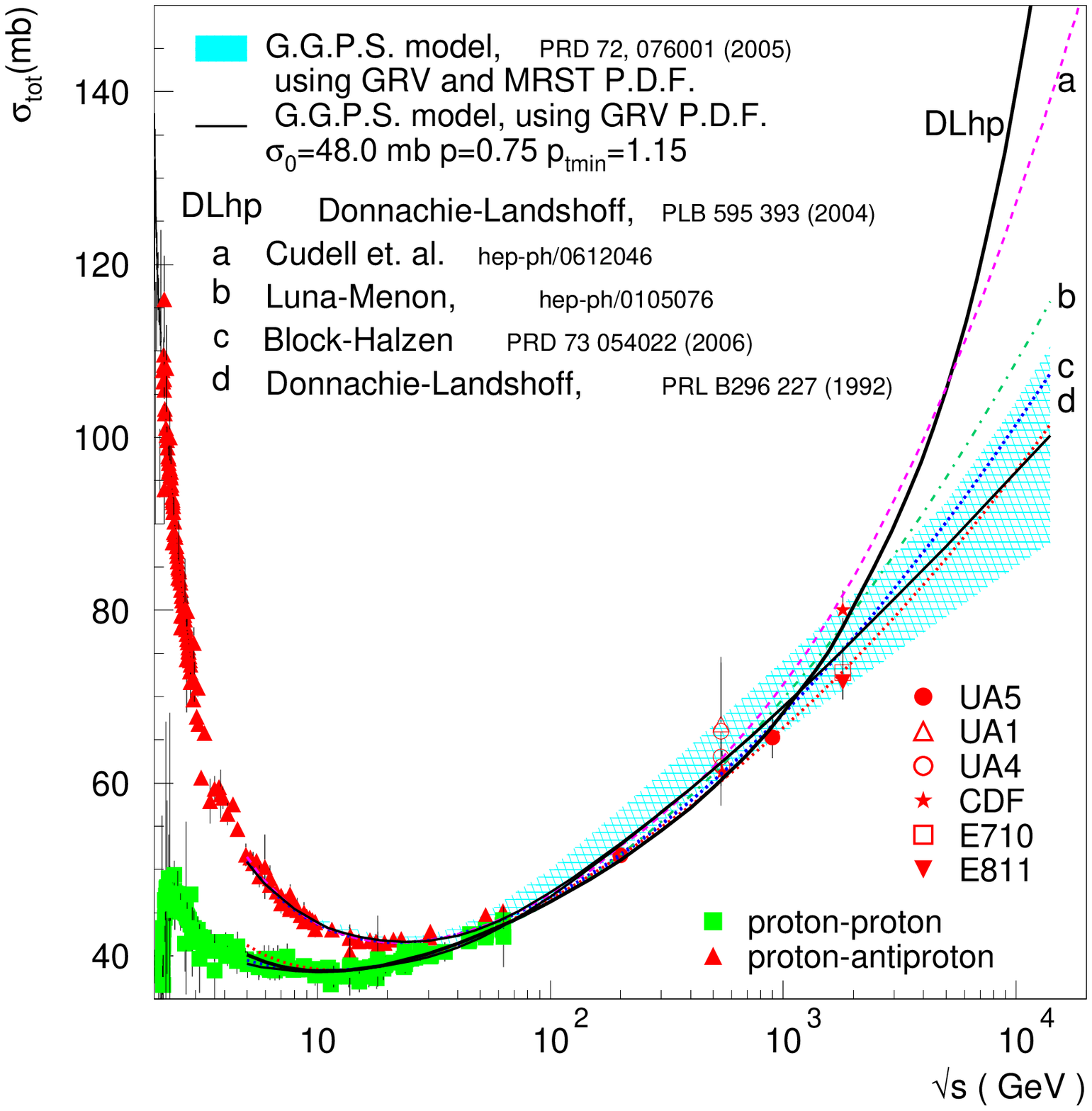}\hspace{0.1cm}
\includegraphics[scale=0.3]{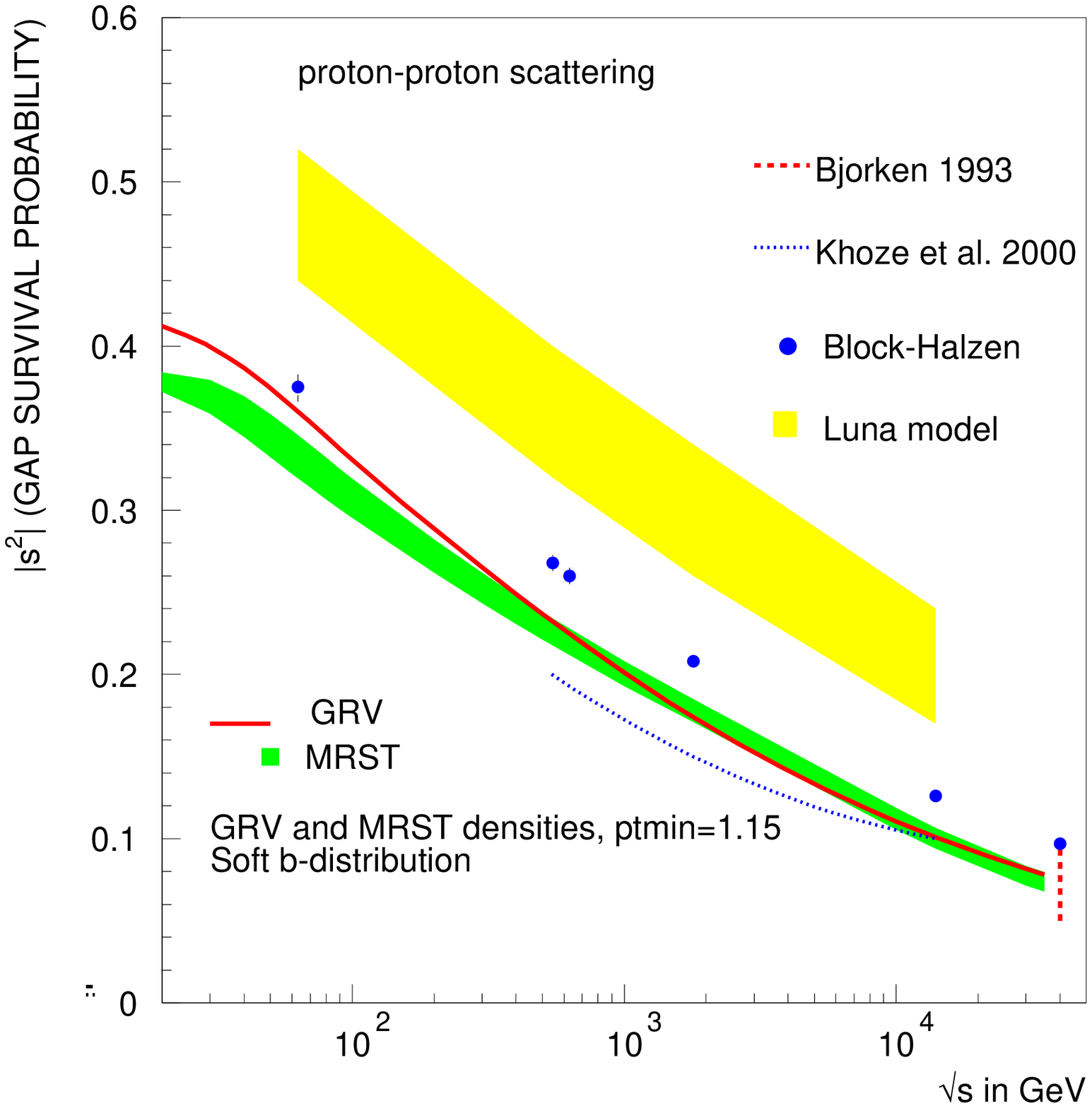}\end{center}
\caption{
Total cross-section data \cite{xdata}  
and models \cite{xmodels}(left) and survival probability for large 
rapidity gaps for different models \cite{survmodels} (right).}
\label{figure1}
\end{figure}

Applying the above described model to the calculation of total 
cross-sections, gives the results shown in the left panel of Figure 
\ref{figure1}. To obtain this figure, we have used different PDF's and 
slightly different values for the parameters $p, \sigma_0$ and 
$p_{tmin}$ and the variations are indicated by the band. We have shown 
\cite{ourplb} that the asymptotic behaviour of this cross-section can be 
fitted with a $\log^2{s}$ type 
%behaviour, confirming a phenomenological satisfation of the Froissart bound. 
behaviour. This is a phenomenological confirmation that the model
satisfies Froissart bound.

We do not see in our model any hard Pomeron behaviour beyond the 
initial rise and predict a value 
$\sigma^{LHC}_{tot}=100^{ +10}_{-13}(mb)$.
The LRGSP can now be calculated using the two quantities 
$e^{-n(b,s)}$ and $A_{BN}^{NP}(b,s)$ which were input in 
the calculation of the total 
cross-section. The right panel of Figure \ref{figure1} shows such evaluations, 
with the yellow band corresponding to various MRST densities 
\cite{MRST}, the central full line to using GRV 
densities \cite{GRV} and the other lines and bands representing comparisons 
with other models, as indicated. The most interesting result of this 
figure is that the predictions for LHC agree reasonably well amongst 
eachother, namely 
$|S|^2=5\div 10\%$, in spite of the fact that the various models differ 
greatly in details and the way in which they achieve results for total 
cross-sections consistent with the Froissart bound.  Thus  the model 
estimates for the LRGSP are quite robust.
\section{Conclusions}
We have built a model for $\sigma_{tot}$ which incorporates hard and  
soft gluon  effects, satisfies the Froissart bound and  can be used 
reliably to study other minimum bias effects e.g. the Survival 
Probability of Large Rapidity Gaps. 
It can also be  extended to calculations of total 
$\gamma p$ and  and $\gamma \gamma$ cross-sections.
\section*{Acknowledgments}
One of us, G.P., wishes to thank the Boston University Theoretical 
Physics Department for hospitality during  the preparation of
this talk.

\end{document}